\documentclass[aps,prd,superscriptaddress,12pt,showpacs,notitlepage]{revtex4}
\usepackage{amsfonts}
\usepackage{graphics}
\usepackage{graphicx}

\newcommand{\be}{\begin{equation}}
\newcommand{\ee}{\end{equation}}
\newcommand{\bea}{\begin{eqnarray}}
\newcommand{\eea}{\end{eqnarray}}
\newcommand{\pa}{\partial}
\newcommand{\bb}{\bibitem}
\newcommand{\dsla}{\rlap/\partial} 
\begin{document}
\title{A simple supersymmetric extension of K field theories}
\author{C. Adam}
\affiliation{Departamento de F\'isica de Part\'iculas, Universidad de Santiago de Compostela and Instituto Galego de F\'isica de Altas Enerxias (IGFAE) E-15782 Santiago de Compostela, Spain}
\author{J.M. Queiruga}
\affiliation{Departamento de F\'isica de Part\'iculas, Universidad de Santiago de Compostela and Instituto Galego de F\'isica de Altas Enerxias (IGFAE) E-15782 Santiago de Compostela, Spain}
\author{J. Sanchez-Guillen}
\affiliation{Departamento de F\'isica de Part\'iculas, Universidad de Santiago de Compostela and Instituto Galego de F\'isica de Altas Enerxias (IGFAE) E-15782 Santiago de Compostela, Spain}


\begin{abstract}
We continue the investigation of  supersymmetric extensions of field theories with a non-standard kinetic term (K field theories) resumed recently.  Concretely, for K field theories which allow for kink or compacton solutions in 1+1 dimensions, i.e., for domain walls in a higher-dimensional context, we find a simple supersymmetric extension such that the boson field still has the kink solution, and the field equation for the fermion in the kink background is linear and is solved by the first spatial derivative of the kink, as is the case in the corresponding standard supersymmetric theories. This supersymmetric extension, nevertheless, is peculiar in several aspects.  The bosonic part of the supersymmetric Lagrangian is {\em not} equal to the original bosonic K field Lagrangian, but the bosonic field equations coincide. Further, the field equation for the bosonic field is produced by the variation of the auxiliary field and vice versa. This observation may be of some independent interest. Finally, the presence of kink solutions does not lead to a central extension in the SUSY algebra, in contrast to the standard case.  
\end{abstract}

\maketitle 

\section{Introduction}

One possible generalization of field theories consists in allowing for non-standard kinetic terms, i.e., kinetic terms which are of higher than second power in field derivatives.  
These theories are known as K field theories. One well-known and rather old example for this type of theories is the Skyrme model \cite{skyrme}, where in addition to the quadratic sigma model term a quartic term is included to allow for the existence of stable static soliton solutions. 
In the realm of ordinary scalar fields the interest in K field theories started to rise more recently, mainly in relation with cosmological applications. Some examples are the so-called K-inflation models as alternative scenarios for cosmological inflation \cite{k-infl}, the K-essence models as alternative explanations of the late time cosmological acceleration \cite{k-ess}, 
\cite{bab-muk-1}, or K field theories as a more general setting for brane world models
\cite{babichev1}-\cite{Dzhu1}. Further, in K field theories some qualitatively new properties may show up, like the existence of solutions with compact support (compactons) \cite{rosenau1}-\cite{bazeia4}.   

Given the possible relevance of K field theories, one rather obvious question is whether they are compatible with supersymmetry, i.e., whether they allow for supersymmetric extensions. 
For the Skyrme--Faddeev--Niemi model (a coset restriction of the Skyrme model to an $S^2$ target space), this issue has been studied already some time ago in \cite{nepo} and, more recently, in \cite{frey}, with the result that the supersymmetric extension requires in any case the addition of further terms to the bosonic part of the Lagrangian; further, it turns out that the auxiliary field becomes dynamical . For scalar field theories which allow for topological kink solutions, i.e., domain walls in higher dimensions, the problem of supersymmetric extensions was investigated quite recently in \cite{bazeia2}, and the present paper intends to further develop on this issue. It can be easily deduced from the results of \cite{bazeia2} that finding the supersymmetric extensions of K field theories is not straight forward. Their extensions require quite nontrivial modifications already of the bosonic part of the extended action, and the field equations for the spinor field typically turn out to be non-linear. Nevertheless, the supersymmetric extensions preserve some properties of the original scalar field theory like, e.g., the existence of topological kink solutions and the corresponding fermionic zero modes.  So one might wonder whether further, simpler supersymmetric extensions of scalar K field theories exist. It is the purpose of our paper to give an affirmative answer to this question and to present and investigate a simple supersymmetric extension of scalar K field theories.   
In Section 2.A, we introduce our conventions, and we briefly discuss the standard scalar supersymmetric kink in Section 2.B for later reference. Then, in Section 2.C, we introduce a different supersymmetric extension of the (non-supersymmetric) scalar kink. This extension is on-shell, i.e., the scalar field equation agrees with the the field equation of the non-supersymmetric scalar theory, whereas the bosonic part of the action is not equal to the action of the non-supersymmetric theory. It is this second supersymmetric extension which can be easily generalized to K theories. In Section 3.A, we briefly describe the  class of scalar K field theories we want to consider, whereas in Section 3.B we introduce the supersymmetric extensions of these K field theories, analogously to what we did in Section 2.C for the standard scalar kink.
In Section 4, we investigate the issue of central extensions of the SUSY algebra in a kink background for our new supersymmetric extensions of K field theories. Finally, Section 5 contains our conclusions.   

\section{Two versions of the supersymmetric kink action}
\subsection{Notation and conventions}
The conventions in \cite{bazeia2} are based on the widely used ones of \cite{Siegel}, so let us first introduce the corresponding notation in our conventions. As in \cite{bazeia2}, we will depart from
the superfield formalism in 2+1 dimensional Minkowski space, because superfields over 1+1 and 2+1 dimensional Minkowski space are very similar, due to the similarity of the spinor representations in the two spaces. We shall restrict to 1+1 dimensions where it is appropriate.
Our only difference with the conventions of \cite{bazeia2} is that our choice of the
Minkowski space metric is $\eta_{\mu\nu} = {\rm diag} (+,-,-)$. All sign differences in some terms in the action between this paper and \cite{bazeia2} can be traced back to this difference.
The scalar superfield is \cite{Siegel}
\be
\Phi(x,\theta)=\phi(x)+\theta^{\alpha}\psi_{\alpha}(x)-\theta^2 F(x)
\ee
where $\phi$ is a real scalar field, $\psi$ is a two-component Majorana spinor, and $F$ is the auxiliary field. Further, $\theta^\alpha$ are the two Grassmann-valued superspace coordinates,
$\psi_\alpha$ is a Grassmann-valued spinor field, and $\theta^2 \equiv (1/2)\theta^\alpha \theta_\alpha$. Spinor indices are risen and lowered with the spinor metric $C_{\alpha \beta} = -C^{\alpha\beta}=(\sigma_2)_{\alpha\beta}$, i.e., $\psi^\alpha = C^{\alpha\beta}\psi_\beta$ 
and $\psi_\alpha = \psi^\beta C_{\beta\alpha}$. 

Next, let us fix the gamma matrix conventions. We want to choose a representation where the components of the Majorana spinor are real. This may be achieved by choosing an imaginary, hermitian $\beta \equiv \gamma^0$ and hermitian, real $\alpha_k \equiv \beta \gamma^k$. Concretely, we choose (the $\sigma_i$ are the Pauli matrices)
\be 
\beta = \sigma_2 \, , \quad \alpha_1 = -\sigma_1 \, , \quad \alpha_2 = - \sigma_3 \quad \Rightarrow \quad \gamma^0 = \sigma_2 \, , \quad \gamma^1 = i\sigma_3 \, , \quad \gamma^2 = -i\sigma_1 .
\ee  
This choice of gamma matrices enables us to introduce the "barred spinor" notation of \cite{shifman}, \cite{shizuya}. The introduction of a second notation may seem a bit artificial, but it turns out that some calculations (especially the rather lengthy ones of section 4) are significantly simpler in this second notation. We define the barred spinor
\be
\bar \psi \equiv \psi^\dagger \gamma^0 = \psi^{\rm T}\gamma^0 \quad \Rightarrow \quad
\bar \psi_\alpha = \psi_\beta (\sigma_2)_{\beta\alpha}.
\ee   
It may be checked easily that the barred spinor is identical to the spinor with upper components in the notation of \cite{Siegel}, $\bar \psi_\alpha \equiv \psi^\alpha =i(\psi_2 ,-\psi_1)$, where $\psi_\alpha = (\psi_1 ,\psi_2 )$. The main advantage of the barred spinor notation is that all spinor indices are lower and we may dispense with the spinor metric.  There is one possible source of confusion related to the use of two different notations, which we resolve by introducing a further bar. The problem is that the gamma matrices should be objects with two lower indices in the barred spinor notation, whereas they should be objects with one lower and one upper spinor index in the spinor metric notation of \cite{Siegel}. That is to say,
\be
(\gamma^\mu \psi)_\alpha \equiv \gamma^\mu{}_\alpha{}^\beta \psi_\beta 
\equiv \bar \gamma^\mu{}_{\alpha\beta}\psi_\beta
\ee
where summation over repeated indices is assumed in both cases.   Here,
\be
\gamma^0{}_\alpha{}^\beta \equiv \bar \gamma^0{}_{\alpha\beta} \equiv (\sigma_2)_{\alpha\beta}
\ee
etc., and obviously $\bar \gamma^\mu{}_{\alpha\beta}$ belongs to the barred spinor notation and should not be confused with $ \gamma^\mu{}_{\alpha\beta} = 
 \gamma^\mu{}_\alpha{}^\gamma C_{\gamma\beta}$. 
 
 In the barred spinor notation, the spinorial expressions assume a simpler and more familiar form, like
 \be
 \bar \chi \psi = \bar \chi_\alpha \psi_\alpha = \chi^\alpha \psi_\alpha \, ,\quad
 \bar \psi \psi = \bar\psi_\alpha \psi_\alpha =\psi^\alpha \psi_\alpha = 2 \psi^2
 \ee
 or
 \be
 \bar \psi \dsla \psi \equiv  \bar \psi \gamma^\mu \pa_\mu \psi = \bar \psi_\alpha 
 \bar \gamma^\mu{}_{\alpha\beta}\partial_\mu \psi_\beta =
\psi^\alpha \gamma^\mu{}_\alpha{}^\beta\partial_\mu \psi_\beta .
\ee
In the barred spinor notation, the scalar superfield reads
\be
\Phi(x,\theta)=\phi(x)+\bar \theta \psi (x)-\frac{1}{2}\bar \theta \theta F(x).
\ee
We remark that we differ in conventions from \cite{shifman}, \cite{shizuya} both in the definition of the $F$ term of the superfield (the $F$ term in the superfield enters with a plus sign in\cite{shifman},  \cite{shizuya}, in contrast to the minus sign employed here and in
\cite{bazeia2}), and in the Grassmann integration convention ($\int d^2 \theta \theta^2 =-1$ here, whereas it is equal to $+1$ in \cite{shifman}, \cite{shizuya}). All sign differences in expressions between the present paper and \cite{shifman}, \cite{shizuya} can be traced back to these different conventions.
Further, observe that there is no factor of $\sqrt{2}$ in the fermion component of the superfield (in contrast to some other conventions),
which will produce some factors of $(1/2)$ in the fermionic part of the component form of the action.  

The components of the superfield can be extracted with the help of the following projections
\be
\label{comp}
\phi(x)=\Phi(z)|,\quad\,\psi_{\alpha}(x)=D_{\alpha}\Phi(z)|,\quad\,F(x)=D^2\Phi(z)|,
\ee
where the superderivative is
\be
D_\alpha = \frac{\partial}{\partial\theta^\alpha} -i \gamma^\mu{}_\alpha{}^\beta
\theta_\beta \partial_\mu = 
\frac{\partial}{\partial \bar\theta_\alpha} -i \bar \gamma^\mu_{\alpha\beta}
\theta_\beta \partial_\mu
\ee
\be
D^2 \equiv \frac{1}{2} D^\alpha D_\alpha = \frac{1}{2}\bar D D
\ee
and the vertical line $|$ denotes evaluation at $\theta^\alpha =0$.

\subsection{The standard supersymmetric kink}
For comparison with later results, let us first briefly review the standard supersymmetric kink theories. 
The simplest scalar superfield action is 
\be
\label{scaf}
S=\int d^3x d^2 \theta \, \left[ - \frac{1}{4}D^{\alpha}\Phi D_{\alpha}\Phi+P(\Phi)\right] =
\int d^3x D^2 \left[-\frac{1}{4}D^{\alpha}\Phi D_{\alpha}\Phi+P(\Phi)\right]\Big| 
\ee
where use was made of the fact that Grassmann integration is equivalent to Grassmann differentiation. Performing the derivatives explicitly and setting $\theta^\alpha$ to zero at the end results in
\be
S=\int d^3 x
\Big[\frac{1}{2}F^2 + \frac{1}{2}i\bar \psi \dsla\psi +
\frac{1}{2}\pa_\mu\phi\pa^\mu \phi+ 
\frac{1}{2}P''(\phi)\bar\psi \psi +P'(\phi)F\Big].
\ee
Finally, eliminating the auxiliary field via its field equation $F=-P'$ we get the standard supersymmetric action
\be
\label{simp}
S=\int d^3 x
\Big[-\frac{1}{2}(P'(\phi))^2 + \frac{1}{2}i\bar\psi \dsla \psi +
\frac{1}{2}\pa_\mu\phi\pa^\mu \phi+\frac{1}{2}P''(\phi)\bar\psi \psi \Big].
\ee
Here, $P(\phi)$ is the prepotential which provides both the potential $V=(1/2)P'^2$ and the Yukawa type interaction $Y=P''$ with the fermion. Observe the presence of the factors $(1/2)$ in the fermionic part of the action in the conventions used here. 

The Euler--Lagrange equations for the action (\ref{simp}) are
\be
\pa_\mu \pa^\mu \phi + V' (\phi) =0
\ee
for the scalar and 
\be
 i\bar \gamma^\mu_{\alpha\beta}\partial_\mu \psi_\beta + Y\psi_\alpha =0
\ee
for the spinor field.
With our gamma matrix conventions we may now specialize the Euler--Lagrange equations to 1+1 dimensions and to the static case of kink or soliton solutions and the corresponding fermionic zero mode equation. Concretely we get 
\be \label{simp-scalar}
\phi_{xx} - V' (\phi)=0 \quad \Rightarrow \quad \frac{1}{2} \phi_x^2 =V
\ee
for the scalar field and
\be \label{simp-fermion}
\mp \partial_x \psi^\pm + Y\psi^\pm =0
\ee
for the fermion field (here $\psi_\alpha = (\psi^+ ,\psi^-)$). If $V$ has more than one vacuum, then there exist finite energy solutions (kinks) of Eq. (\ref{simp-scalar}) which interpolate between different vacua. In the background of such a kink, one of the fermionic zero mode equations (\ref{simp-fermion}) generically has a normalizable solution (e.g. $(\psi^+ ,0)$), whereas the second equation has a non-normalizable solution (e.g. $(0,\psi^-)$). We remark for later use that if we apply, e.g.,  the minus Dirac operator to the plus Dirac (zero mode) equation then we get
\be \label{dirac-pm}
(\pa_x + Y)(-\pa_x + Y)\psi^+ = (-\pa_x^2 +Y' \phi_x + Y^2)\psi^+ = (-\pa_x^2 + V'')\psi^+
\ee
where we used $Y=P''$, $V=(1/2)P'^2$ and $\phi_x = \sqrt{2V}=  P'$. Further, the normalizable solution 
(zero mode) $\psi^+$ is just the derivative of the kink, $\psi^+ = \epsilon \phi_x$ (here $\epsilon $ is a Grassmann-valued constant):
\be
(-\pa_x + Y)\psi^+ = \epsilon (-\pa_x + P'')\phi_x = -\epsilon (\phi_{xx} - P''P') =
-\epsilon ( \phi_{xx} -V') =0.
\ee
This is a consequence of both supersymmetry, which implies that the bosonic and fermionic zero modes (=zero energy solutions of the linear fluctuation equations) in the kink background are the same, and of the translational symmetry of the kink, which implies that the bosonic zero mode is the derivative of the kink. 

\subsection{A new supersymmetric action}

Now let us introduce a new supersymmetric action by simply supersymmetrizing (in the sense of replacing scalar fields by superfields)  the bosonic part of the above action (\ref{simp}). This bosonic part reads 
\be
S_{\rm bos} = \int d^3 x \left(
\frac{1}{2}\pa_\mu\phi\pa^\mu \phi - V(\phi)\right)
\ee
so let us introduce the action
\be \label{superac-f=X}
S=\int d^3x d^2\theta \left(
\frac{1}{2}\pa_\mu\Phi\pa^\mu \Phi - V(\Phi)\right)
\ee
where
\be \label{cal-X}
{\cal X} \equiv \frac{1}{2}\pa_\mu\Phi\pa^\mu \Phi = \frac{1}{2}\pa_\mu\phi\pa^\mu \phi
+ \theta^\alpha \pa_\mu \phi \pa^\mu \psi_\alpha - \theta^2 \left( \pa_\mu \phi \pa^\mu F
+ \frac{1}{2} \pa_\mu \psi^\alpha \pa^\mu \psi_\alpha \right)
\ee
is a genuine superfield like $\Phi$ itself.  In components this action reads
\be \label{f=X-action}
\int d^3 x D^2 ({\cal X} - V(\Phi))|= \int d^3 x \left( \pa_\mu \phi \pa^\mu F + \frac{1}{2} \pa_\mu \psi^\alpha \pa^\mu \psi_\alpha -  \frac{1}{2} V'' (\phi) \psi^\alpha \psi_\alpha -V'(\phi) F\right) .
\ee
In this action, derivatives act on the auxiliary field $F$, so its field equation is no longer algebraic. Nevertheless, this field remains auxiliary in a certain sense, as we shall see in a moment. The field $F$ only appears linearly in the above action, therefore it disappears from its own Euler--Lagrange equation. Indeed, varying w.r.t. $F$ gives the equation
 \be
\pa_\mu \pa^\mu \phi + V' (\phi) =0,
\ee
i.e., the standard field equation of the scalar field.  
In other words, $F$ essentially is a Lagrange multiplier which enforces the standard scalar field equation.

The Euler--Lagrange equation for the fermion field is 
\be
\pa_\mu \pa^\mu \psi_\alpha - V'' \psi_\alpha =0,
\ee
which is not exactly equal to the Dirac equation of the standard theory. However, the two theories share the same zero modes in a kink background, i.e., the same static, one-dimensional solutions. Indeed, the restriction of this equation to one-dimensional, static configurations is identical to Eq. (\ref{dirac-pm}).  

Observe that the auxiliary field $F$ does not show up in the two above equations for $\phi$ and $\psi_\alpha$, i.e., there is no backreaction of $F$ on the evolution of $\phi$ and $\psi_\alpha$. In precisely this sense $F$ still is an auxiliary field. The field $F$ may in principle be calculated from the Euler--Lagrange equation for $\phi$, 
\be
\pa_\mu \pa^\mu F + V'' F + \frac{1}{2}V''' \bar\psi \psi =0 
\ee
once $\phi$ and $\psi$ have been determined,
but due to the auxiliary nature of $F$ in the sense explained above we treat these solutions as physically irrelevant. 

\section{The supersymmetric extended models}

\subsection{K field theories with kinks}

Firstly, let us briefly introduce the K field theories we want to discuss here. The class of bosonic Lagrangians we consider read
\be \label{gen-k}
S_{\rm K, bos} = \int d^3 x \left(
f(X) - V(\phi)\right) , \quad X \equiv \frac{1}{2}\pa_\mu\phi\pa^\mu \phi
\ee
where $f$ is an at the moment arbitrary function of its argument. Several physical conditions (positivity of the energy, global hyperbolicity, well-defined Cauchy problem) may impose further restrictions on $f$. 
The resulting Euler--Lagrange equation is
\be \label{eom-k}
\partial_\mu \left( f'(X) \partial^\mu \phi \right) + V'(\phi )=0.
\ee
For convenience, let us display two explicit examples of these K theories.
A first example is the purely quartic model
\begin{equation}
L= X|X|  - \frac{3}{4} \lambda^2 (\phi^2-1)^2
\end{equation} 
which has the compact kink solutions
\begin{equation}
\phi (x) = \left\{
\begin{array}{lc}
- 1 & x \leq - \frac{\pi}{2 \sqrt{\lambda}} \\
 \sin \sqrt{\lambda}x & \quad
-\frac{\pi}{2\sqrt{\lambda}} \leq x \leq
\frac{\pi}{2\sqrt{\lambda}}  \\
1 & x \geq \frac{\pi}{2\sqrt{\lambda}},
\end{array}
\right. \label{compacton sol}
\end{equation}
see \cite{comp}.
Here the absolute value symbol in the kinetic term is irrelevant for static (compact kink) solutions, but is important to guarantee positivity of the energy in the full, time-dependent system. 

The second example is specifically designed such that the resulting K theory still has the standard $\phi^4$ kink as a solution (of course, the dynamics will be different from the standard $\phi^4$ theory). These K theories have been originally introduced in \cite{bazeia1}, and we just briefly repeat their construction. Indeed, for a general K field theory of the type
(\ref{gen-k}), the field equation for a static, one-dimensional (kink) solution may be integrated once to give
\be \label{gen-kink-eq}
f-2f' X =V
\ee
where $X=-\frac{1}{2}\phi_x^2$ in the static one-dimensional case. For the $\phi^4$ kink $f=X$, $V=\frac{1}{2}(1-\phi^2)^2$, we get the equation $\phi_x^2 = (1-\phi^2)^2$, and the kink solution is $\phi = \tanh x$. For general $f$ we may still assume that  $\phi_x^2 = (1-\phi^2)^2$
(i.e., the standard $\phi^4$ kink solution) and use this condition to determine the corresponding potential in Eq. (\ref{gen-kink-eq}). A specific example of this type is provided by $f=X+\alpha X^2$ where $\alpha$ is a real parameter. Assuming $-2X=\phi_x^2 = (1-\phi^2)^2$ and using Eq. (\ref{gen-kink-eq}) to determine the potential, one gets the Lagrangian density
\be
L=X+\alpha X^2 -\frac{1}{2}(1-\phi^2)^2 -\frac{3}{4}\alpha (1-\phi^2)^4 .
\ee
Other choices for the potential are, of course, possible, but in general they do not lead to closed, analytic expressions for the corresponding kink solutions.

\subsection{The supersymmetric extensions}

In complete analogy with what we did in section 2.B we now supersymmetrize the K field action of the above section in the sense of replacing the scalar field by a superfield.  Doing this, we get the supersymmetric action 
\be \label{superac-f}
S_{\rm K,SUSY}= \int d^3 x d^2 \theta \left(
f({\cal X} )- V(\Phi)\right)
\ee
where ${\cal X}$ is defined in Eq. (\ref{cal-X}). In components this action reads
\bea
S_{\rm K,SUSY} & =& \int d^3 x {\cal L}_{\rm K,SUSY} =
\int d^3 x D^2 \left( f ({\cal X}) - V(\Phi)\right) | = \nonumber \\ &&
\int d^3 x \left[
\frac{1}{2} f''(X) \partial_\mu \phi \partial^\mu \psi^\alpha \partial_\nu \phi \pa^\nu \psi_\alpha 
+ f'(X)\left(
\pa_\mu \phi \pa^\mu F + \frac{1}{2} \pa_\mu \psi^\alpha \pa^\mu \psi_\alpha \right) 
\right.
\nonumber \\ &&
\left.
  -  \frac{1}{2} V'' (\phi) \psi^\alpha \psi_\alpha -V'(\phi) F\right] .
  \label{k-susy-action}
\eea
The field equation for $\phi$ is again provided by the Euler--Lagrange equation w.r.t. the auxiliary field $F$. Explicitly it reads 
\be \label{sc-eom-susy}
\partial_\mu \left( f'(X) \partial^\mu \phi \right) + V'(\phi )=0
\ee
and is, therefore, identical to Eq. (\ref{eom-k}). The Euler--Lagrange equation for the spinor field is 
\be \label{spi-eom-susy}
\pa_\mu \left(f''(X) \pa^\nu \phi \pa_\nu \psi_\alpha \pa^\mu \phi + f'(X) \pa^\mu \psi_\alpha \right) + V''(\phi) \psi_\alpha =0.
\ee
We remark that again the auxiliary field $F$ does not couple to either the scalar or the spinor field and may be treated as auxiliary or unphysical in this sense.  

Finally, let us demonstrate that the fermionic zero mode in a kink background continues to be the derivative of the kink. For a static, one-dimensional scalar field $\phi (x)$ the Euler--Lagrange equation (\ref{sc-eom-susy}) reads
\be
- \pa_x (f'(X)\phi_x) + V'(\phi) =0
\ee
or, with the help of $X_x=-\phi_x \phi_{xx}$, 
\be \label{kink-eq}
f''\phi_x^2 \phi_{xx} - f'\phi_{xx} +V' =0 .
\ee
On the other hand, the Euler--Lagrange equation (\ref{spi-eom-susy}) for a static spinor 
$\psi_\alpha (x)$ in a kink background $\phi(x)$ reads
\be
\pa_x (f'' \phi_x^2 \psi_{\alpha , x} - f' \psi_{\alpha , x })+ V'' \psi_\alpha =0
\ee
and is identically satisfied for a spinor $\psi_\alpha = \epsilon_\alpha \phi_x$ where 
$\epsilon_\alpha $ is a constant spinor. Indeed, inserting this spinor in the above equation we get
\be
\epsilon_\alpha \pa_x (  f''\phi_x^2 \phi_{xx} - f'\phi_{xx} +V') =0
\ee
i.e., just the $x$ derivative of the kink equation (\ref{kink-eq}).

\section{Supercurrent and SUSY algebra}
It is a well-known fact that a standard supersymmetric scalar field theory in 1+1 dimensions has a centrally extended SUSY algebra if it supports topological soliton solutions (kinks) \cite{olive}, where the central charges are related to the topological charges of the solitons. 
Here we want to investigate whether this phenomenon continues to hold  in the case of the supersymmetric extensions of K field theories introduced in the last section. 
The SUSY transformations of the fields are
\bea
\delta \phi = \bar \epsilon \psi &,& \delta \psi = -i\gamma^\mu \epsilon \partial_\mu \phi - \epsilon F \nonumber \\
\delta F = i \bar \epsilon \gamma^\mu \pa_\mu \psi &,&  \delta \bar \psi = i\bar \epsilon \gamma^\mu \pa_\mu \phi - \bar \epsilon F.
\eea
The supersymmetric K field Lagrangian related to the action (\ref{k-susy-action}) transforms under the SUSY transformations by the following total derivative
\be
\delta {\cal L}_{\rm K,SUSY} = i \bar \epsilon \pa_\mu [f'(X) \pa_\nu \phi \gamma^\mu \pa^\nu \psi - V'(\phi) \gamma^\mu \psi ] \equiv \pa_\mu J_2^\mu 
\ee
where the following relations are useful for the calculation,
\be
\bar \epsilon \psi = \bar \psi \epsilon \, , \quad \bar \epsilon \gamma^\mu \psi = - \bar \psi \gamma^\mu \epsilon \, ,
\ee
\be
\bar \epsilon \gamma^\mu \gamma^\nu \psi = \bar \epsilon \left( \frac{1}{2}\{\gamma^\mu , \gamma^\nu \} + \frac{1}{2}[\gamma^\mu ,\gamma^\nu ] \right) \psi = 
\bar \psi \left( \frac{1}{2}\{\gamma^\mu , \gamma^\nu \} - \frac{1}{2}[\gamma^\mu ,\gamma^\nu ] \right) \epsilon \, .
\ee
The part of the SUSY Noether current related directly to the field variations is
\bea
J_1^\mu &\equiv & \left( \delta \phi \frac{\partial}{\pa (\pa_\mu \phi)} +
\delta F \frac{\partial}{\pa (\pa_\mu F)} + \delta \psi \frac{\partial}{\pa (\pa_\mu \psi)} +
\delta \bar\psi \frac{\partial}{\pa (\pa_\mu \bar\psi)} \right) {\cal L}_{\rm K,SUSY} 
\nonumber \\
&=&\bar \epsilon f'(X)[ \pa^\mu F \psi  - F\pa^\mu \psi   + i \pa^\mu \phi \dsla \psi + i \dsla \phi  \pa^\mu \psi  ] \nonumber \\
&+& \bar \epsilon f''(X)\Bigl( \frac{1}{2}(\pa^\mu \bar \psi \pa_\nu \phi \pa^\nu \psi + \pa^\nu \bar \psi
\pa_\nu \phi \pa^\mu \psi )\psi  + \nonumber \\
&& \pa^\mu \phi [(\pa_\nu \phi \pa^\nu F +\frac{1}{2} \pa_\nu \bar\psi \pa^\nu \psi )\psi +
i\pa^\nu \phi \dsla \phi\pa_\nu \psi - F\pa_\nu \phi \pa^\nu \psi ]\Bigr) \nonumber \\
&+& \frac{1}{2}\bar \epsilon f''' (X) \pa^\mu \phi (\pa_\lambda \phi \pa^\lambda \bar \psi \pa_\nu \phi \pa^\nu \psi ) \psi
\eea 
(here in the first line it is understood that the field insertions should be made exactly at the positions where the corresponding field derivatives act), and the full SUSY Noether current is
\be 
J^\mu_{\rm SUSY}= J^\mu_1 - J^\mu_2 \equiv \bar \epsilon {\cal J}^\mu \equiv \bar \epsilon_\alpha {\cal J}^\mu_\alpha
\ee 
where we introduced some notation at the r.h.s.
It may be checked by a lengthy but straight forward calculation that this current is conserved on-shell. 

For an evalution of the SUSY algebra it is useful to study the simpler case $f(X)=X$ (the model of Section 2.C) first. The current in this case is
\be
{\cal J}_\alpha^\mu = \partial^\mu F \psi_\alpha - F \partial^\mu \psi_\alpha + i \partial^\mu \phi
(\dsla \psi)_\alpha + i (\dsla \phi \partial^\mu \psi)_\alpha - i \partial_\nu \phi (\gamma^\mu
\partial^\nu \psi )_\alpha +i V' (\gamma^\mu \psi )_\alpha
\ee
and the correct field equal time (anti) commutators are
\be
[\phi (x), \dot F(y)] = i\delta (x-y) \; , \quad [F(x), \dot \phi (y)] = i \delta (x-y)
\ee
\be
\{ \psi_\alpha (x) , \dot{\bar \psi}_\beta (y)\} = i\delta_{\alpha \beta} \delta (x-y) \; , \quad
\{ \dot \psi_\alpha (x) , \bar \psi_\beta (y)\} = - i\delta_{\alpha \beta} \delta (x-y).
\ee
The bosonic commutators are obvious from the action (\ref{f=X-action}), whereas the anticommutators are obvious up to an overall sign. An easy way to check that our sign choice is right is to observe that with this sign choice the correct SUSY transformations of the fields are produced, i.e., 
\be 
[i\epsilon Q,\phi_n] = \delta \phi_n \qquad \phi_n = (\phi ,\psi, \bar \psi ,F)
\ee
where 
\be
Q_\alpha = \int dx {\cal J}^0_\alpha .
\ee
For the SUSY anticommutator $\{ {\cal J}^0_\alpha (x) ,\bar Q_\beta \}$ we find after another lengthy calculation
\be \label{jq-alg-f=X}
\{ {\cal J}^0_\alpha (x) ,\bar Q_\beta \} = 2 T^0{}_\nu (\bar\gamma^\nu)_{\alpha\beta} + 2i (\bar\gamma^5)_{\alpha\beta} V' \phi '
\ee
(remember $(\gamma^\mu \epsilon)_\alpha \equiv \bar \gamma^\mu{}_{\alpha\beta} \epsilon_\beta \equiv \gamma^\mu{}_\alpha{}^\beta \epsilon_\beta$ in the barred spinor and spinor metric notations, respectively, where $\epsilon$ is an arbitrary spinor; further, $\gamma^5 = \gamma^0 \gamma^1$).
The corresponding energy momentum tensor is
\bea \label{emt-f=X}
T^{\mu\nu} &=& \partial^\mu \phi \partial^\nu F + \partial^\nu \phi \partial^\mu F +
\frac{1}{2} \left( \pa^\mu \bar \psi \pa^\nu \psi + \pa^\nu \bar \psi \pa^\mu \psi \right) - 
\nonumber \\ &&
g^{\mu\nu} \left( \pa_\lambda \phi \pa^\lambda F + \frac{1}{2} \pa_\lambda \bar \psi \pa^\lambda \psi -\frac{1}{2} V'' \bar \psi \psi - V' F\right) .
\eea
It is interesting to contrast this result with the corresponding one for a standard theory like the one in Section 2.B (where the energy-momentum tensor is different, of course), 
\be
\{ {\cal J}^0_\alpha (x) ,\bar Q_\beta \} = 2 T^0{}_\nu (\gamma^\nu)_{\alpha\beta} + 2i (\gamma^5)_{\alpha\beta} P' \phi '.
\ee
The result looks formally almost identical, with the only difference that in the second term at the r.h.s. the prepotential $P$ appears instead of the potential $V$ itself. This difference is, however, important. Indeed, in the standard case a further integration $\int dx$ leads to the SUSY algebra with central extension,
\be 
\{ Q_\alpha , \bar Q_\beta \} = 2 {\cal P}_\nu (\gamma^\nu)_{\alpha \beta} + 2i  
(\gamma^5)_{\alpha\beta} (P(\phi_+) - P(\phi_-))
\ee
where $\phi_\pm = \phi (x=\pm \infty )$, and ${\cal P}_\nu$ is the momentum operator. For a kink, $\phi_+ \not= \phi_-$, and also $P(\phi_+)$ and $P(\phi_-)$ are different, so a central extension appears in the SUSY algebra in a kink background. 

For the anticommutator (\ref{jq-alg-f=X}), on the other hand, a further integral leads to
\be 
\{ Q_\alpha , \bar Q_\beta \} = 2 {\cal P}_\nu (\gamma^\nu)_{\alpha \beta} + 2i  
(\gamma^5)_{\alpha\beta} (V(\phi_+) - V(\phi_-)) = 2 {\cal P}_\nu (\gamma^\nu)_{\alpha \beta} 
\ee
because $\phi_\pm$ must take vacuum values, and $V(\phi)$ is zero by definition for a vacuum value. Therefore, for the theory of Section 2.C there is {\em no} central extension in the SUSY algebra in a kink background. 

It remains to calculate the SUSY algebra for the supersymmetric K field theories of Section 3.B. For this purpose it is useful to introduce the canonical momenta from the variations of the Lagrangian
\bea
\frac{\pa {\cal L}_{\rm K,SUSY}}{\pa (\pa_\mu \phi)} &=& f'''(X)\pa^\mu \phi \pa_\lambda \phi
\pa^\lambda \bar \psi \pa_\nu \phi \pa^\nu \psi + \nonumber \\
&&  f''(X) \left( \pa^\mu \bar \psi \pa_\nu \phi
\pa^\nu \psi + \pa_\nu \phi \pa^\nu \bar \psi \pa^\mu \psi + \pa^\mu \phi \pa_\nu \phi
\pa^\nu F\right) + f'(X) \pa^\mu F
\eea
\be
\frac{\pa {\cal L}_{\rm K,SUSY}}{\pa (\pa_\mu F)} = f'(X) \pa^\mu \phi
\ee
\be
\frac{\pa {\cal L}_{\rm K,SUSY}}{\pa (\pa_\mu \bar \psi_\alpha)} = \frac{1}{2}
\left( f''(X) \pa^\mu \phi \pa_\nu \phi \pa^\nu \psi_\alpha + f'(X) \pa^\mu \psi_\alpha \right)
\ee 
\be
\frac{\pa {\cal L}_{\rm K,SUSY}}{\pa (\pa_\mu  \psi_\alpha)} = \frac{1}{2}
\left( f''(X) \pa^\mu \phi \pa_\nu \phi \pa^\nu \bar \psi_\alpha + f'(X) \pa^\mu \bar \psi_\alpha
\right) .
\ee 
For the bosonic fields we have directly
\be 
\Pi_{\phi} \equiv \frac{\pa {\cal L}_{\rm K,SUSY}}{\pa (\pa_0 \phi)}  \;  , \quad 
\Pi_{F} \equiv \frac{\pa {\cal L}_{\rm K,SUSY}}{\pa (\pa_0 F)} ,
\ee
whereas for the fermi fields we have to take into account that $\psi$ and $\bar \psi$ are not independent, i.e, 
\be
\bar \epsilon_\alpha (\Pi_\psi)_\alpha \equiv (\Pi_{\bar \psi})_\alpha \epsilon_\alpha =
\bar \epsilon_\alpha \frac{\pa {\cal L}_{\rm K,SUSY}}{\pa_0 \bar\psi_\alpha} +  
\frac{\pa {\cal L}_{\rm K,SUSY}}{\pa_0 \psi_\alpha} \epsilon_\alpha
\ee
for an arbitrary spinor $\epsilon$. It follows that e.g.
\be
(\Pi_\psi)_\alpha = f''(X) \dot \phi \pa_\nu \phi \pa^\nu \bar \psi_\alpha + f'(X) \dot {\bar \psi}_\alpha ,
\ee
and the SUSY charge density is
\be
{\cal J}_\alpha^0 =  \psi_\alpha \Pi_\phi + i (\dsla \psi)_\alpha \Pi_f  + 
i (\dsla \phi \Pi_\psi )_\alpha - F (\Pi_\psi )_\alpha - i \pa_\nu \phi (\gamma^0 \pa^\nu \psi)_\alpha 
-i V'(\phi) (\gamma^0 \psi)_\alpha . 
\ee
Finally, the equal time (anti) commutators are
\be
[\phi (x), \Pi_\phi (y)] = i\delta (x-y) \; , \quad [F(x), \Pi_F (y)] = i \delta (x-y)
\ee
\be
\{ \psi_\alpha (x) , (\Pi_\psi)_\beta (y)\} = i\delta_{\alpha \beta} \delta (x-y) \; , \quad
\{ \bar \psi_\alpha (x) , (\Pi_{\bar \psi})_\beta (y)\} = - i\delta_{\alpha \beta} \delta (x-y)
\ee
(the anticommutators for $\psi$ and $\bar \psi$ are of course not independent). For the SUSY charge and charge density algebra we find again Eq. (\ref{jq-alg-f=X}).
The SUSY algebra in a kink background, therefore, again contains {\em no} central extension.
The energy-momentum tensor is, of course,  different from the one in Eq. (\ref{emt-f=X}). Its explicit expression is rather long and not particularly illuminating, therefore we do not display it here.

\section{Summary}
We explicitly constructed a simple supersymmetric extension of scalar field theories with a non-standard kinetic term (K field theories). This supersymmetric extension is of the on-shell type,
that is, the field equations of the original bosonic K field theory coincide with the bosonic field equations of the supersymmetric extension. The bosonic part of the supersymmetric Lagrangian, on the other hand, does not coincide with the original K field Lagrangian. 
Also the role played by the auxiliary field $F$ is different from the standard case, although it remains true that the auxiliary field does not influence the dynamics of either the bosonic or the fermionic fields. 

Concretely, we considered K field theories which give rise to (compact or non-compact) topological kink solutions in 1+1 dimensions, which correspond to domain walls in a higher-dimensional context. As in the case of standard SUSY kink theories, also in the non-standard supersymmetric extension each kink solution supports a fermionic zero mode. In contrast to the standard case, however, the non-standard supersymmetric extension in a kink background does {\em not} lead to a central extension of the SUSY algebra.

Comparing our results with the corresponding ones of \cite{bazeia2}, we remark that the supersymmetric extensions considered in \cite{bazeia2} do not reproduce the standard scalar kink equation (\ref{eom-k}) for the scalar field. Instead, already the scalar field equations are more complicated, although in the static case they still may be reduced to first order equations, and provide solutions with a kink-like or compacton shape. That is to say, the relation between the original scalar field theories and its supersymmetric extensions  is more involved in the cases studied in \cite{bazeia2}. 

The following observations make the results presented in the present paper especially interesting.  Firstly,  domain walls are possible sources of structure formation in the early universe. Further, in many theoretical considerations (e.g. in scalar field models derived from or motivated by string theory), supersymmetry is supposed to be unbroken at high energies. In this context, supersymmetric extensions of field theories which support defect structures are, therefore,  of direct interest  for cosmological investigations of the early universe. 
We want to remark that in cosmological applications the physically relevant information is provided by the structure of solutions (e.g., defect structures or energy densities driving inflation), therefore an {\em on-shell} supersymmetric extension is completely natural from this point of view.  
Secondly, our method of constructing the supersymmetric extensions is rather generic (see e.g. Eqs. (\ref{superac-f=X}), (\ref{superac-f})). Therefore, it should not be too difficult to generalize this method to higher dimensions, like e.g. the Skyrme or baby Skyrme models. Here, the generalization to 2+1 dimensions will be more straight forward (because the SUSY algebra is essentially the same), whereas the case of 3+1 dimensions will require some modifications. In any case, we think we have proposed a new, simple way of constructing supersymmetric extensions of non-linear field theories with generalized kinetic terms which has not been explored so far and which deserves further investigation.       

\section*{Acknowledgement}
The authors acknowledge financial support from the Ministry of Science and Investigation, Spain (grant FPA2008-01177), and
the Xunta de Galicia (grant INCITE09.296.035PR and
Conselleria de Educacion). They thank R. Nepomechie for useful comments. 
Further, JMQ thanks A. Ramallo for helpful suggestions and discussions.

\end{document}